\documentclass[runningheads]{llncs}
\usepackage{hyperref}
\usepackage{orcidlink}
\usepackage{graphicx}
\usepackage{marvosym} 
\usepackage{algorithmic}
\usepackage{amsmath} 
\usepackage{amssymb}
\usepackage{booktabs}
\usepackage{enumitem}
\usepackage{colortbl}
\usepackage{xcolor}
\usepackage{subcaption} 
\usepackage{multirow}
\usepackage{bigdelim}
\usepackage{arydshln}
\usepackage{diagbox} 
\usepackage{wrapfig}
\usepackage{wasysym}
\usepackage{makecell}

\usepackage[linesnumbered,ruled,vlined]{algorithm2e}
\SetKwInput{KwInput}{Input}    
\SetKwInput{KwOutput}{Output}    
\SetKwInput{KwFunction}{Function}

\renewcommand{\footnotesize}{\small}

\usepackage{tikz}
\usepackage{graphicx}
\usepackage{xcolor}
\newcommand{\circleone}[1]{%
    \resizebox{!}{0.8em}{%
        \tikz[baseline=(char.base)]{
            \node[shape=circle, fill=black, inner sep=0.8pt, text=white] (char) {#1};
        }%
    }%
}

\newcommand{\circletwo}[1]{%
    \resizebox{!}{0.8em}{%
        \tikz[baseline=(char.base)]{
            \node[shape=circle, fill=black, inner sep=0.8pt, text=white] (char) {#1};
        }%
    }%
}

\newcommand{\circlethree}[1]{%
    \resizebox{!}{0.8em}{%
        \tikz[baseline=(char.base)]{
            \node[shape=circle, fill=black, inner sep=0.8pt, text=white] (char) {#1};
        }%
    }%
}

\usepackage[capitalize]{cleveref}
\crefname{section}{Sec.}{Secs.}
\Crefname{section}{Section}{Sections}
\Crefname{table}{Table}{Tables}
\crefname{table}{Tab.}{Tabs.}
\DeclareCaptionStyle{ruled}{labelfont=normalfont,labelsep=colon,strut=off}

\definecolor{gray}{rgb}{0.78,0.7843,0.792}
\def\ie{\textit{i.e.}}
 
\def\eg{\textit{e.g.}}
\def\etal{\textit{et al.}}
\def\aka{\textit{a.k.a.}}
\makeatletter

\newcommand{\Rmnum}[1]{\expandafter\@slowromancap\romannumeral #1@}
\makeatother

\newcommand{\ct}[1]{\texttt{#1}}

\usepackage{xcolor}
\definecolor{sotacolor}{RGB}{152,52,48}
\definecolor{runupcolor}{RGB}{236,84,40}

\usepackage[absolute,overlay]{textpos}
\begin{document}
\begin{textblock}{13}(1.5,1)
\centering
\textit{This paper has been accepted by ESORICS 2024}
\end{textblock} 
    \title{ECLIPSE: Expunging Clean-label Indiscriminate Poisons via Sparse Diffusion Purification}
 
\titlerunning{Expunging Poisons via Sparse Diffusion Purification}

\author{
Xianlong Wang\inst{1,2,4,5 \dag}\orcidlink{0009-0009-3057-827X}
\and
Shengshan Hu\inst{1,2,4,5 \dag \text{(\Letter)}}\orcidlink{0000-0003-0042-9045} 
\and 
Yechao Zhang\inst{1,2,4,5 \dag}\orcidlink{0000-0002-0551-1200}
\and
Ziqi Zhou\inst{1,2,3 \S}\orcidlink{0009-0000-6785-7306}
\and
Leo Yu Zhang \inst{\dag\dag} \orcidlink{0000-0001-9330-2662}  
\and Peng Xu \inst{1,2,4,5 \dag \text{(\Letter)}} \orcidlink{0000-0003-4268-4976}
\and 
Wei Wan\inst{1,2,4,5 \dag} \orcidlink{0000-0002-1247-5092}
\and
Hai Jin \inst{1,2,3 \S} \orcidlink{0000-0002-3934-7605}
}

\authorrunning{X. Wang et al.} 

\institute{
\textsuperscript{1}National Engineering Research Center for Big Data Technology and System\\
\textsuperscript{2}Services Computing Technology and System Lab \
\textsuperscript{3}Cluster and Grid Computing Lab \\
\textsuperscript{4}Hubei Engineering Research Center on Big Data Security \\ \textsuperscript{5}Hubei Key Laboratory of Distributed System Security \\ 
\textsuperscript{\dag}School of Cyber Science and Engineering, Huazhong University of Science and Technology, Wuhan 430074, China\\
\textsuperscript{\S}School of Computer Science and Technology, 
Huazhong University of Science and Technology, Wuhan 430074, China \\
\textsuperscript{\dag\dag}School of Information and Communication Technology, Griffith University, Southport QLD 4215, Australia
\email{\{wxl99,hushengshan,ycz,zhouziqi,xupeng,wanwei\_0303,hjin\}@hust.edu.cn}
\email{leo.zhang@griffith.edu.au} 
}

\maketitle               
\begin{abstract}
Clean-label indiscriminate poisoning attacks add invisible perturbations to correctly labeled training images, thus dramatically reducing the generalization capability of the victim models. Recently, defense mechanisms such as adversarial training, image transformation techniques, and image purification have been proposed. However, these schemes are either susceptible to adaptive attacks, built on unrealistic assumptions, or only effective against specific poison types, limiting their universal applicability. In this research, we propose a more universally effective, practical, and robust defense scheme called ECLIPSE. We first investigate the impact of Gaussian noise on the poisons and theoretically prove that any kind of poison will be largely assimilated when imposing sufficient random noise. In light of this, we assume the victim has access to an extremely limited number of clean images (\textit{a more practical scene}) and subsequently enlarge this sparse set for training a denoising probabilistic model (\textit{a universal denoising tool}). We then introduce Gaussian noise to absorb the poisons and apply the model for denoising, resulting in a roughly purified dataset. Finally, to address the trade-off of the inconsistency in the assimilation sensitivity of different poisons by Gaussian noise, we propose a lightweight corruption  compensation module to effectively eliminate residual poisons, providing a more universal defense approach. Extensive experiments demonstrate that our defense approach outperforms 10 state-of-the-art defenses. We also propose an adaptive attack against ECLIPSE and verify the robustness of our defense scheme. 
Our code is available at \url{https://github.com/CGCL-codes/ECLIPSE}.

\end{abstract}

\section{Introduction}
\label{sec:intro}

The success of \textit{deep neural networks} (DNNs) relies on abundant training data, motivating many commercial firms to supply their training set by automatically scraping images from untrusted sources. 
However, these untrusted data have the potential to be exploited by adversaries to poison DNNs, challenging their trustworthiness in safety-critical applications~\cite{fowl2021preventing}.

Recently, there has been a rise in the occurrence of clean-label indiscriminate poisoning attacks that add imperceptible perturbations to correctly labeled images, thus dramatically compromising DNNs. These perturbations are usually norm-bounded and together with clean labels, constitute the concealment of such attacks, making them easier to implement in real-world scenarios.
Based on this, 
in this research, we focus on  \textit{clean-label indiscriminate poisoning attacks with bounded perturbations} (CLBPAs)~\cite{chen2022self,fowl2021preventing,adv,fu2021robust,unl,sandoval2022autoregressive,tensorclog,wen2023adversarial,syn,yuan2021neural},  
which introduce a great challenge for defenders.

\begin{figure}[t]
\centering
\scalebox{0.87}{  
\includegraphics[width=\textwidth]{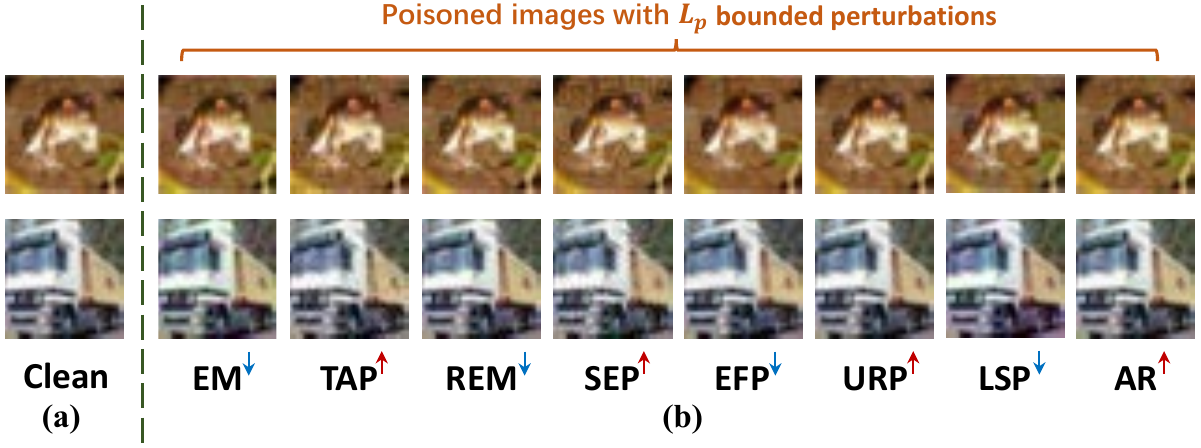}}
\caption{We present eight popular clean-label indiscriminate poisoning attacks along with clean samples. The upward and downward arrows represent high-frequency and low-frequency poison perturbations, respectively. It can be observed that it is difficult for the naked eyes to distinguish between clean samples and poisoned samples.}
\label{fig:2} 
\end{figure}

Existing defense strategies have been successively proposed  but suffer from the following limitations:   
\circleone{1}\textbf{ Limited effectiveness against certain CLBPA types.} 
Many defense schemes are only effective against specific types of CLBPAs, \eg, the grayscale transformation in  \textit{image shortcut squeezing} (ISS)~\cite{liu2023image} is only effective against low-frequency poisons, and OP~\cite{sandoval2023can} only works when facing class-wise poisons. 
It is crucial to design a more universally applicable defense against CLBPAs since the concealment of poisons makes it difficult for defenders to identify the type of bounded perturbations being used as shown in Fig.~\ref{fig:2} (b); 
\circletwo{2}\textbf{ Making impractical assumptions.} 
Several purification schemes~\cite{dolatabadi2023devil,jiang2023unlearnable} are proposed to defend against CLBPAs via diffusion denoising. However, these approaches are impractical as they make unrealistic assumptions about the clean training set, \eg, Dolatabadi~\etal~\cite{dolatabadi2023devil} assume the defender can obtain the whole clean training set to train a diffusion model, which seriously violates the assumption of CLBPA implemented during the training phase;  
\circlethree{3}\textbf{ Fragile to adaptive attacks.} Many vulnerable defense schemes are easily compromised by adaptive attacks shortly after their proposal, \eg, Tao~\etal~\cite{bettersafe} suggest that \textit{adversarial training} (AT) can address CLBPAs, but a series of adaptive attacks~\cite{fu2021robust,wen2023adversarial} subsequently  compromise AT. ISS is also susceptible to adaptive attacks, as acknowledged by~\cite{liu2023image}.

Additionally, it is intricate to determine whether the training set is clean or poisoned due to the concealment of bounded poisons as shown in~\cref{fig:2} (a) and (b). As a result, any defense against CLBPAs must be applied without significantly compromising accuracy in the absence of poisons. 
Based on this, we propose clean accuracy, \ie, 
model accuracy when applied with defense on the clean dataset, to serve as another significant evaluation metric that has been underrepresented in prior works. 
We then reassess previous \textit{state-of-the-art} (SOTA) defenses~\cite{liu2023image,bettersafe} with this metric in~\cref{tab3} and find that they both negatively impact clean training to some extent.
Therefore, there is an urgent need to design a defense scheme against CLBPAs that is \textit{more universally effective,  practical, robust to adaptive attacks, and does not substantially impair clean accuracy.}

Our intuition starts with a key observation (\cref{fig3} (a)) and a theoretical guarantee (Theorem 1), \ie, the morphology of various poisons can be assimilated by gradually introducing random Gaussian noise. 
Once by denoising the noised input, we can effectively eliminate the noise as well as poison, which motivates us to apply a denoising diffusion probabilistic model~\cite{diffusion} to handle the denoising process. 
In the context of CLBPA where acquiring ample clean training data is not feasible, the challenge of training a diffusion model emerges as a new obstacle for us. We are constrained to assume the defender has privately stored an extremely limited amount of data (less than 5\% of the size of the training dataset from the main task) that is distributed identically to the clean training set. To overcome data scarcity in our practical assumption, we propose leveraging a data enlargement module to augment the dataset used for training the diffusion model. 

Additionally, owing to discrepancies in the assimilation effects caused by diverse poison patterns, some poisons necessitate more noise for assimilation.  
However, this will also lead to excessive noise absorption for other poisons, resulting in a damage of image features. To address the trade-off between the purification effects of diverse poisons, after introducing moderate Gaussian noise and applying the denoising process, we further propose a lightweight compensation module that incorporates both probabilistic grayscale transformation and the lightweight Gaussian noise. Only in this way can we achieve a more universally effective and practical defense strategy against diverse CLBPAs, while demonstrating superior \textit{clean accuracy} compared to existing SOTA defense approaches. 
Furthermore, we devise an adaptive CLBPA based on information gleaned from our defense, thereby showcasing the effectiveness of our defense in countering this attack.
Our main contributions can be summarized as follows:

\begin{itemize}[label=\textbf{\phantom{$\bullet$}\textbullet}]
\item We establish the theoretical and experimental evidence supporting the assimilation of poisons by  Gaussian noise.
Leveraging this insight, we propose a more universally effective defense using forward noise addition and a denoising process facilitated by the diffusion model.

\item We point out the existing diffusion purification schemes for defending against CLBPAs are impractical as they require a large amount of clean training samples. Instead, we propose training the diffusion model solely with sparse, identically distributed clean data.

\item To address the trade-off of the inconsistency in the assimilation sensitivity of different poisons by Gaussian noise, we propose a lightweight compensation module to remove residual poisons and provide an explanation of the roles of corruptions from a frequency perspective.

\item Extensive experiments on multiple benchmark datasets including CIFAR-10 and ImageNet demonstrate that our defense scheme can outperform the SOTA defense methods in test accuracy by 38.4\% on CIFAR-10 using ResNet18 and in clean accuracy by 4.41\% on CIFAR-10. Additionally, our defense's effectiveness remains robust against our newly proposed adaptive poisoning attack, further affirming its reliability. 

\end{itemize}

\section{Related Work}

\subsection{Clean-label Indiscriminate Poisoning Attacks}
Traditional indiscriminate poisoning attacks inject noisy labels~\cite{biggio2011support,munoz2017towards,zhang2017game} and can be easily detected by human observers. 
So existing works focus on clean-label indiscriminate poisoning attacks~\cite{adv,sandoval2022autoregressive,wen2023adversarial}, \aka, clean-label generalization attack~\cite{feng2019learning,yuan2021neural}, clean-label availability poisoning attack~\cite{syn}, delusive attack~\cite{bettersafe} or simply unlearnable examples~\cite{unl}, which contaminate correctly labeled training images with $\mathcal{L}_p$ norm bounded perturbations, \ie, CLBPAs, to ensure stealthiness, including solving bi-level optimization problems to produce \textit{error-minimizing noises} (EM)~\cite{unl} or serving \textit{targeted adversarial samples} (TAP) as poison perturbations~\cite{adv}. Chen~\etal~\cite{chen2022self} enhanced the poisoning effectiveness by employing \textit{self-ensemble of model checkpoints} (SEP). 
In addition, some adaptive CLBPAs designed for AT have also been proposed one after another, \eg, REM~\cite{fu2021robust} and EFP~\cite{wen2023adversarial}, reducing the defense universality of AT.
While the above schemes often rely on external networks, resulting in substantial time costs, several model-agnostic CLBPAs have emerged.
For instance, \textit{universal random perturbation} (URP) adds the same random Gaussian noise to images from the same category, offering a simpler and more efficient poisoning attack~\cite{unl,bettersafe}.  
\textit{Linearly separable perturbations} (LSP)~\cite{syn} and \textit{autoregressive poisons} (AR)~\cite{sandoval2022autoregressive} also fall into this category, prioritizing efficiency and transferability.


\subsection{Defenses Against Poisoning Attacks}
There are many approaches available to defend against data poisoning attacks, including  differential privacy \cite{hong2020effectiveness}, and strong data augmentations \cite{borgnia2021strong,cutout,cutmix,mixup}. 
However, these schemes are not specifically optimized to handle CLBPAs and proved significantly less effective in addressing them based on our experimental results. In light of this, Tao~\etal~\cite{bettersafe} experimentally and theoretically proved that AT~\cite{pgd,zhou2023advclip,zhou2023downstream,zhou2024securely} can be applied to defend against CLBPAs. 
Unfortunately, some stronger adaptive CLBPAs against AT are proposed and can effectively break AT~\cite{fu2021robust,wang2021fooling,wen2023adversarial}. 
Subsequently, Liu~\etal~\cite{liu2023image} found that grayscale transformation is effective against low-frequency poisons and JPEG compression is effective against high-frequency poisons, which can effectively defend against CLBPAs. But its fatal flaw is that each simple transformation only has the best defense effect for specific types of poisons, lacking a universal solution. 
Soon after, Qin~\etal~\cite{qin2023learning} introduced adversarial augmentation, Sandoval~\etal~\cite{sandoval2023can} proposed orthogonal projection, but they both only work against certain poisons and are not a universally effective defense solution. 
Another defensive route involves image purification~\cite{dolatabadi2023devil,jiang2023unlearnable} through
using a diffusion model to denoise the poisoned samples. But they made the unrealistic assumption of owning ample clean training images, seriously violating the scenario definition of data poisoning attacks.  
It is desirable but challenging to design a versatile, practical, and robust defense approach against CLBPAs.


\section{Methodology}

\subsection{Threat Model}
\label{sec3.1}
Following the standard framework of CLBPAs~\cite{chen2022self,adv,unl,ren2022transferable,Sandoval_2022_CVPR,bettersafe,wen2023adversarial,syn}, we assume the attacker manipulates all the training images with bounded perturbations with $L_p$ norm. The attacker aims to cause the model $G$ with parameter $\theta$ trained on the poisoned dataset to generalize poorly to a clean data distribution $\mathcal{D}$. Formally, the attacker expects to work out the following bi-level objective:
\begin{equation} 
\label{eq1}
\centering
\max  \underset{(x, y) \sim \mathcal{D}}{\mathbb{E}}\left[\mathcal{L}\left(G\left(x ; \theta_{p}\right), y\right)\right]
\end{equation}
\begin{equation}
\label{eq2}
 \text { s.t. } \theta_{p} =  \underset{\theta}{\arg \min } \sum_{\left(x_{i}, y_{i}\right) \in  \mathcal{D}_{c}} \mathcal{L}\left(G\left(x_{i}+\delta_i ; \theta\right), y_{i}\right)
\end{equation}
where $(x_{i}, y_{i})$ represents the clean data belonging to the clean training set $\mathcal{D}_{c}$, $\delta_i$ is the elaborate perturbation to poison the training set with $L_p$ norm constraint, and $\mathcal{L}$ is a loss function, \eg, cross-entropy loss.
As for defenders, we only assume that they access to an extremely low proportion, \eg, 5\% of clean samples from the same training  distribution. The defenders aim to perform operations on the poisoned images to achieve the opposite goal of~\cref{eq1}, while not involving any knowledge of the victim models.

\subsection{Motivation for Studying Defenses Against CLBPAs}
CLBPAs inject malicious noise into training data, causing a decline in the performance of DNN models, which poses significant harm in real-world scenarios. For instance, poisoning data collected by web crawlers during the training of large models can degrade model performance~\cite{adv,sandoval2022autoregressive,wen2023adversarial}. Additionally, poisoning internal data used for peer assessment or academic research by proxy applications and subsequently training models on these contaminated data can result in severely degraded performance~\cite{feng2019learning}. Therefore, researching defense mechanisms against CLBPAs holds strong practical significance for a wide range of 
DNN-based technologies or applications in real-world settings.

\subsection{Key Intuition and Theoretical Insight}

\begin{figure*}[t]
  \centering
  \scalebox{1}{  
  \includegraphics[width=\textwidth]{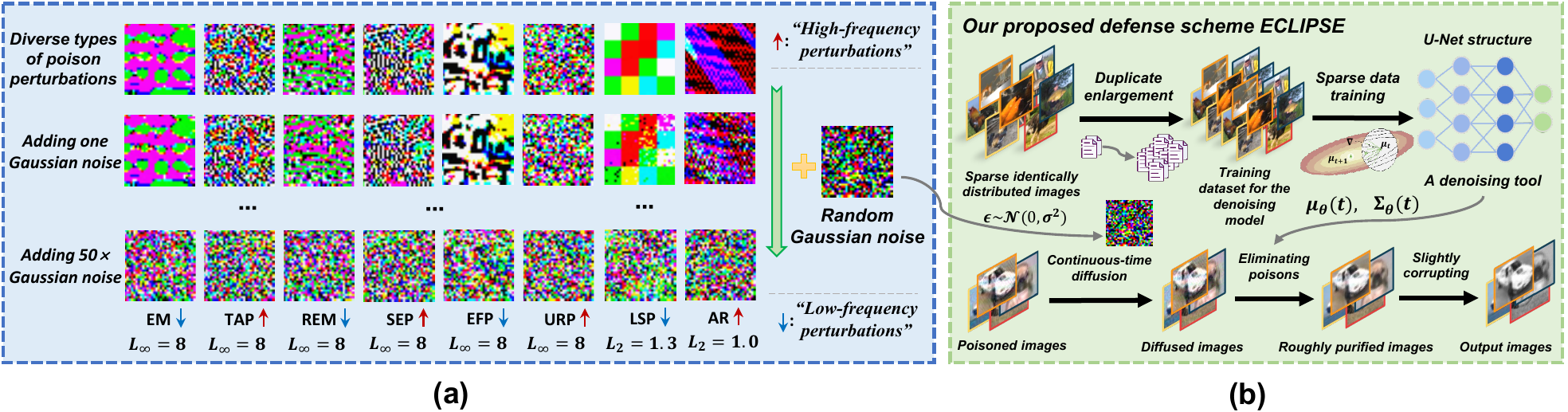}}
  \caption{(a) We present eight types of poison perturbations and add Gaussian noise that is subject to normal distribution $\mathcal{N}(0, 0.01^2)$ from one to fifty rounds gradually. We observe that the assimilation of Gaussian noise to low-frequency perturbations is slow, while the assimilation to high-frequency perturbations is faster; (b) The high-level overview of our proposed defense scheme ECLIPSE.}
  \label{fig3} 
\end{figure*}

We visually distinguish eight  poison patterns as illustrated in the top row of Fig.~\ref{fig3} (a). 
These distinct perturbation patterns underscore the complexity of mitigating CLBPAs with a universal scheme, which is completely different from expunging adversarial perturbations composed of only one high-frequency pattern via the existing diffusion purification scheme~\cite{diffpure}. 
Therefore, this poses a thought-provoking question for us: 
\begin{quote}
    \centering
    \emph{Can diffusion purification expunge both high and low-frequency poison perturbations?}
\end{quote}
To answer this, we progressively apply incremental random Gaussian noise to these poisons, then the patterns from different poisoning attacks begin to resemble each other, ultimately converging to Gaussian noise, as shown in Fig.~\ref{fig3} (a).
This observation suggests that bounded poison perturbations can be ultimately assimilated by random Gaussian noise. 
Regardless of how the images themselves change, it will not affect the actual assimilation effect of the Gaussian noise we add.
Additionally, we provide the theoretical insight for this conclusion:

\noindent\textbf{Theorem 1: }
\textit{Assuming $p(x, t)$ and $q(x, t)$ represent the poisoned data distribution $p(x, \cdot)$ and clean data distribution $q(x, \cdot)$ after undergoing the forward Gaussian noise process with time $t$, respectively (note that the poison perturbations in $p(x, \cdot)$ are constrained within an $L_p$ norm ball), we have:}
\begin{equation*}
    \frac{\partial  D_{KL}\left(p(x, t) \| q(x, t)\right)}{\partial t} \le 0
\end{equation*}
where $D_{KL}$ denotes Kullback-Leibler divergence. 

\noindent\textbf{Proof:} \textit{See Appendix.} 

\noindent This theorem indicates that as the noise level in the forward process increases, the distribution of the poisoned dataset becomes closer to the distribution of the clean dataset after adding noise, which means that the impact of any poison diminishes over time, \ie, \textit{the continuously added noise will eventually absorb all types of bounded poison perturbations.} 
Thus we propose serving the diffusion model as a more  universally 
effective denoising tool for eliminating diverse types of poison perturbations.

\subsection{Challenges and Approaches}
To design a more universally effective defense strategy against existing CLBPAs, we suffer from several challenges as follows:

\textbf{Challenge \Rmnum{1}: In the context of data poisoning attacks, the absence of the clean training set prevents the training of a diffusion model.} 
Existing diffusion defenses are unrealistically assumed by Dolatabadi~\etal~\cite{dolatabadi2023devil} to obtain 100\% clean training images to train a diffusion model and Jiang~\etal~\cite{jiang2023unlearnable} to  obtain 20\% clean training images to fine-tune a clean data trained diffusion model.
These impractical assumptions motivate us to propose a more realistic one.  
Firstly, we assume the defender \textit{only owns sparse identically distributed clean data instead of directly owning any clean training images.} 
Secondly, our defender \textit{does not require any pre-trained diffusion models to fine-tune, thus training from scratch using sparse data.}

\textbf{Challenge \Rmnum{2}:  The universal Gaussian noise scale will lead to asynchrony in the absorption of different poisons.} 
From Fig.~\ref{fig3} (a), it can be observed that low-frequency poisons are assimilated more slowly, indicating that less Gaussian noise is beneficial for absorbing high-frequency poisons while low-frequency poisons cannot be completely absorbed. 
On the other hand, more Gaussian noise is advantageous for low-frequency poisons but may negatively impact the features of images containing high-frequency poisons. 

The promising approach to resolve this dilemma is to set an appropriate value of the intensity of added noise, which is sufficient to absorb high-frequency poison perturbations 
and then design a compensation module to further expunge the residual poisons while simultaneously minimizing harm to purified  poisoned images as much as possible.

\subsection{Our Design for ECLIPSE}
The high-level overview of our defense approach ECLIPSE is shown in~\cref{fig3} (b) and the specific implementation steps are as follows.

\begin{figure}[t]
\centering
\scalebox{0.55}{  
\includegraphics[width=\textwidth]{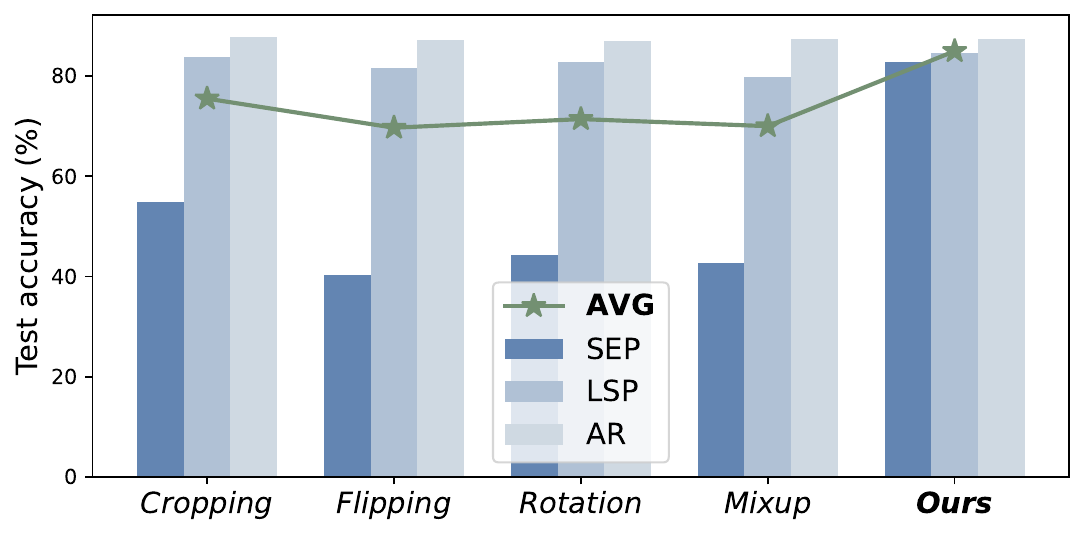}}
\caption{The defense performance of ECLIPSE using diverse data augmentation techniques and our scheme against three CLBPAs, SEP~\cite{chen2022self}, AR~\cite{sandoval2022autoregressive}, and LSP~\cite{syn} using ResNet18 on the CIFAR-10 dataset}
\label{fig:aug} 
\end{figure}

\noindent\textbf{Sparsely training a denoising tool.}
We assume that the defender privately stored a sparse image set $\mathcal{D}_s = \{s_i\}_{i=1}^B$, which \textit{only shares the same distribution as the clean training set (the size is $N$)}. 
Ensuring that the sparse dataset and the distribution of clean data are from the same distribution is crucial~\cite{dolatabadi2023devil,diffpure}.
We set $B \ll N$ ($\frac{B}{N}$ is less than 5\%), making our assumption more practical in poisoning attacks. 
To address the trade-off between practical assumption and the size of sparse set, we attempt to augment the dataset using various standard data augmentation techniques, including \textit{cropping}, \textit{flipping}, \textit{rotation}, and the strong data augmentation  \textit{mixup}~\cite{mixup}. 
Unfortunately, the use of these data augmentations has proven ineffective in defending against certain attacks,~\eg, SEP~\cite{chen2022self} (see~\cref{fig:aug}), limiting the universal effectiveness of our defense solution. 
We speculate that this is because the essence of training diffusion models lies in the learning the mapping from the noised data distribution to the clean data distribution, and yet the data augmentations alter the original clean data distribution~\cite{zhang2024does}, impacting the sampling ability of diffusion models.

To address this, we propose to directly duplicate the original dataset, thus maintaining the distribution of the augmented dataset entirely consistent with the original, while also increasing the volume of data (the results in~\cref{fig:aug} demonstrate the effectiveness of our augmenting approach for training diffusion models with sparse data). We formulize our repetitive data enlargement scheme as:
\begin{equation}
    \mathcal{D}_A = \mathcal{D}_s \cup \mathcal{R}(\mathcal{D}_s, M)
\end{equation}
where function $\mathcal{R}$ denotes the dataset obtained after performing the replication operation on the dataset $\mathcal{D}_s$, $M$ represents the number of replications, and $\mathcal{D}_A$ represents the enlarged dataset used to train the diffusion model.
  
We then employ an unconditional diffusion process to generate $x_1, x_2, \ldots, x_T$ based on an initial image $x_0$ sampled from our enlarged dataset $\mathcal{D}_A$. The forward random Gaussian noise-adding process is formulated as:
\begin{equation}
\label{eq4}
    q\left(x_{0}, x_{1}, \ldots, x_{T} \right)=q\left(x_{0}\right)\prod_{t=1}^{T} q\left(x_{t} \mid x_{t-1}\right)
\end{equation}
\begin{equation}
\label{eq5}
    q\left(x_{t} \mid x_{t-1}\right)=\mathcal{N}\left(x_{t} ; \alpha_{t} x_{t-1}, \beta_{t} \mathbf{I}\right)
\end{equation}
where $q\left(x_{0}, x_{1}, \ldots, x_{T} \right)$ is the joint distribution of forward process, $\beta_{t}$ is the variance of random noise at time $t$, $\alpha_{t}$ and $\beta_{t}$ satisfy $\alpha_{t} ^{2} + \beta_{t} ^{2} = 1 $. 
The training optimization goal is:
\begin{equation}
\label{eq6}
    \min D_{KL}(q \| p)=\int 
    q\log \frac{q}{p} d {x}_{0} d {x}_{1} \cdots d {x}_{T}
\end{equation}
where $p$ represents the joint distribution of estimated reverse process, $D_{KL}$ denotes the  Kullback-Leibler divergence, which is 
used to measure the similarity between two distributions. By simplifying Eq.~(\ref{eq6}), ignoring the constant obtained from the integration and reducing coefficients as suggested by \cite{diffusion}, the loss function becomes:
\begin{equation}
\label{eq7}
L_s = \mathbb{E}\left[\left\|\boldsymbol{\epsilon}-\boldsymbol{\epsilon}_{\boldsymbol{\theta}}\left(\sqrt{1-\bar{\alpha}_{t}} \boldsymbol{\epsilon}+\sqrt{\bar{\alpha}_{t}} {x}_{0}, t\right)\right\|^{2}\right]
\end{equation}
where $\bar{\alpha}_{t}=\alpha_{1} \cdot \alpha_{2} \cdots \alpha_{t}$, $\boldsymbol{\epsilon} \sim \mathcal{N}(\mathbf{0}, \boldsymbol{I})$, ${x}_{0} \sim {q} \left({x}_{0}\right)$, $\boldsymbol{\epsilon}_{\boldsymbol{\theta}}$ is used to predict $\boldsymbol{\epsilon}$ from $x_t$ working as a function approximator. We utilize a cosine-based variance schedule, set larger diffusion steps, and predict a mixing vector $v$ to learn diagonal variance as an interpolation between $\tilde{\beta}_t=\beta_{t}\cdot\left ( {1-\bar{\alpha}_{t-1}} \right ) /\left (  {1-\bar{\alpha}_{t}}  \right ) $ and $\beta_{t}$  to achieve a better training process of diffusion as suggested by Nichol and Dhariwal~\cite{ddim}. 
Thus the new optimization objective is:
\begin{equation}
\label{eq8}
\begin{aligned}
L &= L_s + \gamma \sum_{t=0}^{T} L_t, 
L_{0}  :=-\log p_{\theta}\left(x_{0} \mid x_{1}\right) \\
L_{t-1}  :=D_{KL}(q (x_{t-1} & \mid  x_{t}, x_{0}) \| p_{\theta}(x_{t-1} \mid x_{t}) ), 
L_{T} :=D_{KL}(q(x_{T} \mid x_{0}) \| p\left(x_{T}\right))
\end{aligned}
\end{equation}
where $p_{\theta}(x_{t-1} \mid x_{t})=\mathcal{N}(x_{t-1};\mu_{\theta}\left(x_{t}, t\right), \Sigma_{\theta}\left(x_{t}, t\right))$, $\gamma$ is a hyper-parameter. The learnable denoising parameters
$\mu_{\theta}(x_{t},t)$ and $\Sigma_{\theta}\left(x_{t}, t\right)$ are calculated as:
\begin{equation}
\label{eq:mu}
    \mu_{\theta}(x_{t},t)=\frac{x_t-\frac{1-\alpha _t}{\sqrt{1-\bar{\alpha} _t } } \boldsymbol{\epsilon}_{\boldsymbol{\theta}}(x_{t} , t)}{\sqrt{\alpha _t} }  
\end{equation}
\begin{equation}
\label{eq:sigma}
    \Sigma_{\theta}\left(x_{t}, t\right)=\exp((1-v)\log\tilde{\beta}_t + v\log\beta_t )
\end{equation}

\begin{algorithm}[t]
\SetAlgoLined
\DontPrintSemicolon
\caption{Our defense scheme ECLIPSE}
\SetKwInOut{Input}{Input}
\Input {Poisoned dataset $\{ ({x_p}_i,y_i)\mid i=1,2,...,N \}$; sparse image set $\mathcal{D}_{s} = \{ s_i\mid i=1,2,...,B  \}$ ($B\ll N$); replication times $M$;  training iteration $I$; diffusion step $T$; forward step $t^*$; grayscale probability $p$; standard  deviation $\sigma$}      
\SetKwInOut{Output}{Output}
\Output {Final dataset $ \{ ({x_f}_i,y_i)\mid i=1,2,...,N \}$}
\SetKwInOut{Function}{Function}
\Function{Loss $L$; $\alpha \left ( t \right )$; corruption function $C\left(\boldsymbol{\cdot} ; p,\sigma\right)$.}

Initialize $\mathcal{D}_{A} = \mathcal{D}_{s}$;

\For{$i = 1$ to $M$}{
    $\mathcal{D}_A = \mathcal{D}_A \cup \mathcal{D}_s $; 
    \quad \textcolor{blue} {$\triangleright$ enlarge the sparse set}
}

\For{$i = 1$ to $I$}{
    $\boldsymbol{\epsilon} \sim \mathcal{N}(0, I)$, $t \sim U(1,T)$;
    
    Randomly sample image $x_0$ from $\mathcal{D}_{A}$;
    
    Perform a gradient descent step on $ \nabla_\theta L$; \quad \textcolor{blue} {$\triangleright$ train the diffusion model}
}

Obtain a diffusion model with $\mu_{\theta}$ and $\Sigma_{\theta}$;

\For{$i=1$ to $N$}{
    $\boldsymbol{\epsilon} \sim \mathcal{N}(0, I)$;
    
    Noised image $x_{t^{*}}=\sqrt{\alpha\left(t^{*}\right)} \boldsymbol{x}_{p_i}+\sqrt{1-\alpha\left(t^{*}\right)} \boldsymbol{\epsilon}$;
    
    \For{$t=t^*$ to $1$}{
        \If{$t > 1$}{
            $\boldsymbol{z} \sim \mathcal{N}(0, I)$;
        }
            
        \Else{
            $\boldsymbol{z} = \boldsymbol{0}$;
        }
            
        $x_{t-1}=\mu_{\theta}(x_{t},t) + \Sigma_{\theta}\left(x_{t}, t\right) \mathbf{z}$;  \quad\textcolor{blue} {$\triangleright$ image denoising}
    }

    Receive a largely purified image ${x_e}_i$;
    
    ${x_f}_i = C({x_e}_i; p, \sigma)$;
}
\textbf{Return:} Final dataset $\{ ({x_f}_i,y_i)\mid i=1,2,...,N \}$.
\label{alg1}
\end{algorithm}

\noindent\textbf{Absorbing and eliminating diverse poisons.}
Motivated by our key intuition and theoretical insight, we first add random Gaussian noise to poisoned images for absorbing the poison perturbations, which is implemented by the continuous-time diffusion mode \cite{song2021scorebased} as:
\begin{equation}
\label{eq11}
x_{t^{*}}=\sqrt{\alpha\left(t^{*}\right)} {x}_{p}+\sqrt{1-\alpha\left(t^{*}\right)} \boldsymbol{\epsilon}
\end{equation}
where $\alpha \left ( t \right ) =e^{k_{1}t^2+k_{2}t} $, $k_{1}$ and $k_{2}$ are constants below 0, $\boldsymbol{\epsilon} \sim \mathcal{N}(0, I)$, $x_p$ is the poisoned image, forward step $t^*$ represents the strength of Gaussian noise added. 
After absorbing process, we employ the denoising parameters in~\cref{eq:mu,eq:sigma} from the sparse data trained diffusion model to eliminate poison perturbations, which is defined as:
\begin{equation}
\label{eq12}
    x_{t-1}=\mu_{\theta}(x_{t},t) + \Sigma_{\theta}\left(x_{t}, t\right)\mathbf{z}
\end{equation}
where $\mathbf{z} \sim \mathcal{N} (0, I) $ and  we set $\mathbf{z}=\mathbf{0}$ when $t=1$.

\noindent\textbf{Lightweight corruption compensation module.} 
Owing to variations in the assimilation effects induced by different poison patterns, certain poisons require a greater amount of noise for effective assimilation.  
Specifically, we observe that low-frequency poisons (\eg, EM, REM, and LSP) and robust high-frequency poisons (\eg, SEP) are assimilated more slowly as suggested in Tab.~\ref{tab4} and analyzed in Sec.~\ref{sec:ablation}.  
To address this, we propose a lightweight corruption compensation module to expunge these residual poison perturbations while ensuring that image features are not excessively harmed.  
Since the low-frequency poison operates in color-sensitive regions of the image, we utilize the \textit{probabilistic grayscale transformation} to remove residual low-frequency poisons. 
In addition, we first propose the \textit{lightweight Gaussian noise} to eliminate robust high-frequency poison, \ie, SEP (see~\cref{sec:analysis}). 
The two-stage lightweight corruption techniques are both capable of effectively expunging residual poisons while ensuring minimal impact on image features, which can be formally defined as:
\begin{equation}
\label{eq13}
    x_f = C(x_e; p, \sigma) = \begin{cases}
 G(x_e)+\boldsymbol{\varepsilon}  & \text{with probability } p \\
 x_e+\boldsymbol{\varepsilon}  & \text{with probability } 1-p
\end{cases}
\end{equation}
where $x_f$ represents the final processed image, $x_e$ denotes the purified image, $\boldsymbol{\varepsilon} \sim \mathcal{N}(0, \sigma^2)$, and $G$ represents the grayscale transformation function. 
Please refer to the Algorithm 1 for the detailed process of ECLIPSE.

 \section{Experiments}

\subsection{Experimental Settings}
\textbf{Implementation details.}
The forward timestep $t^*$ is 100, $M$ is 4, training iteration $I$ is 250$K$, grayscale probability $p$ is 0.4, and standard deviation $\sigma$ is 0.05 unless otherwise stated. We use 4\% images with the same distribution as the training set of CIFAR-10~\cite{cifar10}, 1.5\% images with the same distribution as the training set of ImageNet~\cite{imagenet} to serve as sparse sets.  
Diverse network structures including ResNet~\cite{resnet}, VGG~\cite{vgg}, and DenseNet~\cite{densenet} are selected. 
We use SGD for training with a momentum of 0.9, a learning rate of 0.1, and a batch size of 128 for 80 epochs.

\vspace{-5mm}
\begin{table*}[h]
\centering
\scalebox{0.55}
{
\begin{tabular}{c|cccccccccc|ccccccccc}
\toprule[1.5pt] 
\cellcolor[rgb]{.95,.95,.95}\textbf{Models $\rightarrow$} & \multicolumn{10}{c|}{\cellcolor[rgb]{.95,.95,.95}\textbf{ResNet18~\cite{resnet}}}  & \multicolumn{9}{c}{\cellcolor[rgb]{.95,.95,.95}\textbf{VGG19~\cite{vgg}}}  \\
\cellcolor[rgb]{.95,.95,.95}\textbf{Defenses$\downarrow\quad$ Attacks$\rightarrow$}  & \cellcolor[rgb]{.95,.95,.95}EM    & \cellcolor[rgb]{.95,.95,.95}TAP   & \cellcolor[rgb]{.95,.95,.95}REM   & \cellcolor[rgb]{.95,.95,.95}SEP   & \cellcolor[rgb]{.95,.95,.95}EFP   & \cellcolor[rgb]{.95,.95,.95}URP   & \cellcolor[rgb]{.95,.95,.95}LSP   & \cellcolor[rgb]{.95,.95,.95}AR  &\cellcolor[rgb]{.95,.95,.95}ADP  & \cellcolor[rgb]{.95,.95,.95}\textbf{AVG}   & \cellcolor[rgb]{.95,.95,.95}EM    & \cellcolor[rgb]{.95,.95,.95}TAP   & \cellcolor[rgb]{.95,.95,.95}REM   & \cellcolor[rgb]{.95,.95,.95}EFP   & \cellcolor[rgb]{.95,.95,.95}URP   & \cellcolor[rgb]{.95,.95,.95}LSP   & \cellcolor[rgb]{.95,.95,.95}AR  &\cellcolor[rgb]{.95,.95,.95}ADP   & \cellcolor[rgb]{.95,.95,.95}\textbf{AVG}   \\ 

\midrule[1.5pt]

w/o     & 17.58 & 26.16 & 27.34 & 9.01  & 86.64 & 16.80 & 24.48 & 10.59 & 25.93	&27.17   & 19.90 & 27.81 & 31.22 & 81.83 & 16.53 & 21.80 & 13.94 &24.02 &29.63    \\ 

\ldelim\{{5}{*}[\hspace{-2mm}\clap{\footnotesize 
\parbox{2cm}{\centering \hspace{-12mm}\textit{Invalid} \\ \hspace{-12mm}\textit{defenses}}}] 
\colorbox[RGB]{243,243,243}{Cutout~\cite{cutout}}  & 17.63 & 29.16 & 22.42 & 9.30  &  87.76 &  84.41 & 22.93 & 12.92 &21.39 	&\cellcolor[rgb]{.953,.953,.953}34.21  & 39.51 & 31.69 & 22.87 & 85.06 & 29.20 & 25.92 & 10.44 &18.38  &\cellcolor[rgb]{.953,.953,.953}32.88  \\

\colorbox[RGB]{243,243,243}{Mixup~\cite{mixup}}   
& 30.74 & 24.36 & 29.99 & 8.35  & 88.76 & 17.29 & 23.48 & 11.46 &31.20 &\cellcolor[rgb]{.953,.953,.953}29.51   & 22.75 & 28.38 & 33.30 & 82.49 & 16.32 & 24.59 & 14.47 &37.14 &\cellcolor[rgb]{.953,.953,.953}32.43    \\

\colorbox[RGB]{243,243,243}{Cutmix~\cite{cutmix}}  & 29.73 & 23.70 & 29.42 & 6.66  &  87.74 &  84.24 & 20.86 & 14.56 &23.14 	&\cellcolor[rgb]{.953,.953,.953}35.56 & 29.85 & 25.09 & 30.26 &  81.71 & 10.14 & 25.27 & 15.71 &24.16 	&\cellcolor[rgb]{.953,.953,.953}30.27  \\

\hspace{3.5mm}\colorbox[RGB]{243,243,243}{DP-SGD~\cite{hong2020effectiveness}}  & 18.17 & 30.96 & 25.92 & 8.24  &  87.98 & 20.49 & 22.11 & 10.19 &21.51 &\cellcolor[rgb]{.953,.953,.953}27.29  & 20.45 & 27.09 & 24.52 & 84.23 & 30.77 & 25.73 & 14.33 &27.66 &\cellcolor[rgb]{.953,.953,.953}31.85  \\ \hdashline

\ldelim\{{5}{*}[\hspace{-1mm}\clap{\footnotesize 
\parbox{2cm}{\centering \hspace{-10mm}\textit{Limited} \\ \hspace{-10mm}\textit{validity}}}]

\colorbox[RGB]{254,250,224}{ISS-G~\cite{liu2023image}}    & 88.42 & \cellcolor[rgb]{.996,.98,.878}21.88 & 65.29 & \cellcolor[rgb]{.996,.98,.878}7.57  &  86.54 & 60.64 & 65.89 & \cellcolor[rgb]{.996,.98,.878}38.64 &\cellcolor[rgb]{.996,.98,.878}34.42 &52.14  & 86.47 & \cellcolor[rgb]{.996,.98,.878}27.57 & 71.16 & 83.23 & 60.84 &  80.91 & \cellcolor[rgb]{.996,.98,.878}39.60 &\cellcolor[rgb]{.996,.98,.878}41.60 &61.42  \\

\colorbox[RGB]{254,250,224}{AA~\cite{qin2023learning}}  & 85.30 	&67.12 	&\cellcolor[rgb]{.996,.98,.878}39.73 	&\cellcolor[rgb]{.996,.98,.878}24.94 	& 87.76 	& 90.81 	& 87.38    	&51.19 	&58.22 	&65.83 &78.99 	&56.81 	&\cellcolor[rgb]{.996,.98,.878}10.00 	&78.71 	& 82.73 	&\cellcolor[rgb]{.996,.98,.878}9.99 	&\cellcolor[rgb]{.996,.98,.878}25.32 	&\cellcolor[rgb]{.996,.98,.878}24.35 	&\cellcolor[rgb]{.996,.98,.878}45.86 \\

\colorbox[RGB]{254,250,224}{OP~\cite{sandoval2023can}}  &65.42 	&\cellcolor[rgb]{.996,.98,.878}45.86 	&\cellcolor[rgb]{.996,.98,.878}30.44 	&\cellcolor[rgb]{.996,.98,.878}10.01 	& 82.64 	& 89.28 	& 90.14 &\cellcolor[rgb]{.996,.98,.878}33.60 	&\cellcolor[rgb]{.996,.98,.878}33.80 	&53.47  	&79.79 	&54.09 	&\cellcolor[rgb]{.996,.98,.878}31.88 	&78.65 	& 87.14  & 87.43 	&\cellcolor[rgb]{.996,.98,.878}13.64 	&\cellcolor[rgb]{.996,.98,.878}29.47 	&57.76    \\

\hspace{5.5mm}\colorbox[RGB]{254,250,224}{AVATAR~\cite{dolatabadi2023devil}} &\cellcolor[rgb]{.996,.98,.878}27.45 	& 86.63 	&\cellcolor[rgb]{.996,.98,.878}35.74 	&\cellcolor[rgb]{.996,.98,.878}44.97 	&75.90 	& 86.86 	&\cellcolor[rgb]{.996,.98,.878}39.93 	& 83.98 	&67.95 	&61.05 &\cellcolor[rgb]{.996,.98,.878}34.92 	& 83.63 	&\cellcolor[rgb]{.996,.98,.878}39.41 	&74.57 	& 84.08 	&53.96 	& 83.49 &65.22 	&64.91   	 \\  \hdashline

\ldelim\{{3}{*}[\hspace{-4mm}\clap{\footnotesize 
\parbox{2cm}{\centering \hspace{-12mm}\textit{Qualified} \\ \hspace{-12mm}\textit{defenses}}}]

\colorbox[RGB]{220,242,241}{AT~\cite{bettersafe}}      & 68.31 & \cellcolor[rgb]{.863,.949,.945}82.46 & 60.80 & 63.23 & 71.46 & \cellcolor[rgb]{.863,.949,.945}85.63 & \cellcolor[rgb]{.863,.949,.945}81.94 &\cellcolor[rgb]{.863,.949,.945}84.04 &\cellcolor[rgb]{.863,.949,.945}\textbf{82.76}	&75.63  & 64.31 &\cellcolor[rgb]{.863,.949,.945}81.33 & 63.28 & 66.63 &\cellcolor[rgb]{.863,.949,.945}83.22 & 79.15 & \cellcolor[rgb]{.863,.949,.945}80.69 &\cellcolor[rgb]{.863,.949,.945}\textbf{80.58} 	&74.90  \\

\colorbox[RGB]{220,242,241}{ISS-J~\cite{liu2023image}}   & 78.35 & \cellcolor[rgb]{.863,.949,.945}80.77 &\cellcolor[rgb]{.863,.949,.945}81.54 &\cellcolor[rgb]{.863,.949,.945}80.93 & 70.54 & \cellcolor[rgb]{.863,.949,.945}81.27 & 79.55 & \cellcolor[rgb]{.863,.949,.945}81.39 &\cellcolor[rgb]{.863,.949,.945}80.98	&79.48  & 78.69 & \cellcolor[rgb]{.863,.949,.945}80.93 & 79.16 & 68.75 & \cellcolor[rgb]{.863,.949,.945}80.74 & 78.49 &\cellcolor[rgb]{.863,.949,.945}81.56 &78.46  &78.35 \\
 
\colorbox[RGB]{220,242,241}{\textbf{ECLIPSE (Ours)}} &\cellcolor[rgb]{.863,.949,.945}\textbf{82.80} 	&\cellcolor[rgb]{.863,.949,.945}\textbf{86.13} 	&\cellcolor[rgb]{.863,.949,.945}\textbf{82.72} 	&\cellcolor[rgb]{.863,.949,.945}\textbf{82.85} 	&\textbf{77.20} 	&\cellcolor[rgb]{.863,.949,.945}\textbf{86.98} 	&\cellcolor[rgb]{.863,.949,.945}\textbf{84.58} 	&\cellcolor[rgb]{.863,.949,.945}\textbf{87.32} 	&\cellcolor[rgb]{.863,.949,.945}82.62	&\cellcolor[rgb]{.863,.949,.945}\textbf{83.69} 	&\cellcolor[rgb]{.863,.949,.945}\textbf{80.73} 	&\cellcolor[rgb]{.863,.949,.945}\textbf{84.83} 	&\textbf{79.90} 	&\textbf{75.87} 	&\cellcolor[rgb]{.863,.949,.945}\textbf{85.86} 	&\cellcolor[rgb]{.863,.949,.945}\textbf{83.48} 	&\cellcolor[rgb]{.863,.949,.945}\textbf{85.84} 	&\cellcolor[rgb]{.863,.949,.945}80.11 	&\cellcolor[rgb]{.863,.949,.945}\textbf{82.08}   \\ \bottomrule[1.5pt]
\end{tabular}  
}
\caption{\textbf{Main results:} The \textit{test accuracy} (\%) results on CIFAR-10 with ResNet18 and VGG19.  ``\textbf{AVG}" denotes the average value of each row, ``ADP" denotes our proposed adaptive attack against ECLIPSE. The \textbf{bold values} denote the best defense effect among the qualified defense schemes.}
\label{tab:tab1} 
\end{table*}
\vspace{-10mm}

\subsection{Evaluation of ECLIPSE}
\noindent\textbf{Comparison baselines.}
We compare with five SOTA defenses, \ct{ISS}~\cite{liu2023image}, \ct{OP}~\cite{sandoval2023can}, 
\ct{AA}~\cite{qin2023learning}, \ct{AVATAR}~\cite{dolatabadi2023devil}, and \ct{AT}~\cite{bettersafe}. Besides, other common defenses such as \ct{DP-SGD} \cite{hong2020effectiveness,zhang2021feddpgan}, \ct{cutmix} \cite{cutmix}, \ct{mixup} \cite{mixup}, and \ct{cutout} \cite{cutout} are tested. 

\noindent\textbf{Evaluation metrics.} 
Two evaluation metrics are used to evaluate these defense schemes: (i) \textbf{\textit{test accuracy}}, \ie, the accuracy of the model obtained after applying the defense against CLBPAs on clean test set, and (ii) \textbf{\textit{clean accuracy}}, \ie, the accuracy of the model obtained after applying the defense against the clean training set on clean test set.

\noindent\textbf{Main results.}  
The values of average test accuracy that are similar between post-defense and undefended scenarios, are covered by \colorbox[RGB]{243,243,243}{gray} demonstrated in~\cref{tab:tab1}. 
This indicates that~\ct{cutout},~\ct{mixup},~\ct{cutmix}, and~\ct{DP-SGD} are almost ineffective for CLBPAs. 
We also highlight the results with \colorbox[RGB]{254,250,224}{accuracy below 50\% in light yellow} to denote the unqualified defense and \colorbox[RGB]{220,242,241}{accuracy above 80\% in light blue} to indicate that the defense capability is considered excellent. 
Therefore, \ct{ISS-G}, \ct{AA}, \ct{OP}, and \ct{AVATAR} exhibit extreme limitations in countering various types of poisons as shown in~\cref{tab:tab1}, rendering them unsuitable as universal defense solutions. 
As also demonstrated in~\cref{tab:tab1}, two SOTA defense schemes~\ct{AT} and \ct{ISS-J}, also lag behind ECLIPSE by more than 8\% and 4\% in average test accuracy, respectively. 
In addition, our defense also outperforms these two SOTA defense solutions on ImageNet as shown in~\cref{tab:tab2}.

Given that only \ct{AT} and \ct{ISS-J} achieve comparable defense performance in test accuracy, we further only compare the clean accuracy of these two defenses in~\cref{tab3}. It can be seen that ECLIPSE has an absolute and significant advantage in this metric. 
Meanwhile, \ct{ISS-J} \textit{causes a damage of approximately 9\% on clean training, constituting a fatal flaw that compromises this approach} (the values in~\cref{tab:tab2,tab3} covered by \colorbox[RGB]{244,176,132}{deep orange} denote the optimal defense effect, while \colorbox[RGB]{248,203,173}{light orange} denotes the suboptimal).

\begin{table}[t]
\centering
\scalebox{0.86}
{
\begin{tabular}{c|>{\centering\arraybackslash}m{0.9cm}
>{\centering\arraybackslash}m{0.9cm}>{\centering\arraybackslash}m{0.9cm}>{\centering\arraybackslash}m{1.2cm}>{\centering\arraybackslash}m{0.9cm}|>{\centering\arraybackslash}m{0.9cm}>{\centering\arraybackslash}m{0.9cm}>{\centering\arraybackslash}m{0.9cm}>{\centering\arraybackslash}m{1.2cm}>{\centering\arraybackslash}m{0.9cm}}
\toprule[1.5pt]
\cellcolor[rgb]{.95,.95,.95}\textbf{Architectures}   & \multicolumn{5}{c|}{\cellcolor[rgb]{.95,.95,.95}\textbf{ResNet18}}   & \multicolumn{5}{c}{\cellcolor[rgb]{.95,.95,.95}\textbf{DenseNet121}}          
\\ \cellcolor[rgb]{.95,.95,.95}\textbf{Defenses$\downarrow$ Poisons$\rightarrow$} & \cellcolor[rgb]{.95,.95,.95}TAP  & \cellcolor[rgb]{.95,.95,.95}URP  & \cellcolor[rgb]{.95,.95,.95}AR   & \cellcolor[rgb]{.95,.95,.95}CLEAN & \cellcolor[rgb]{.95,.95,.95}\textbf{AVG}
& \cellcolor[rgb]{.95,.95,.95}TAP & \cellcolor[rgb]{.95,.95,.95}URP  & \cellcolor[rgb]{.95,.95,.95}AR   & \cellcolor[rgb]{.95,.95,.95}CLEAN & \cellcolor[rgb]{.95,.95,.95}\textbf{AVG} \\ 
\midrule[1.5pt]
w/o & 40.8 & 51.3 & 25.7 & 72.1  & 47.5    &46.2     & 69.5 & 22.4 & 77.8  & 54.0 \\  
ISS-G & 28.5 & 27.8 & 19.5 & 56.9  & 33.2    &31.2     & 40.5 & 23.7 & 62.7  & 39.5 \\
AVATAR & 52.3 & \cellcolor[rgb]{.957,.69,.518}65.3 & 54.0 & \cellcolor[rgb]{.957,.69,.518}72.3  & 61.0    &57.0    & \cellcolor[rgb]{.957,.69,.518}68.1 & 48.9 & \cellcolor[rgb]{.957,.69,.518}73.1  & 61.8  \\
AT & 58.2 & \cellcolor[rgb]{.973,.796,.678}61.9 & 47.3 & 66.8  & 58.5  & \cellcolor[rgb]{.957,.69,.518}61.8       & 63.1 & 41.9 & 71.9  & 59.7        \\
ISS-J & \cellcolor[rgb]{.973,.796,.678}61.2 & 60.8 & \cellcolor[rgb]{.957,.69,.518}59.4 & 66.2  & \cellcolor[rgb]{.973,.796,.678}61.9 & \cellcolor[rgb]{.973,.796,.678}61.6
& 60.8 & \cellcolor[rgb]{.957,.69,.518}62.0 & 68.6  & \cellcolor[rgb]{.973,.796,.678}63.3         \\
\textbf{ECLIPSE (Ours)}  & \cellcolor[rgb]{.957,.69,.518}61.9 & 60.4 & \cellcolor[rgb]{.973,.796,.678}58.9 & \cellcolor[rgb]{.973,.796,.678}67.0  & \cellcolor[rgb]{.957,.69,.518}\textbf{62.1}     &60.9
& \cellcolor[rgb]{.973,.796,.678}63.4 &\cellcolor[rgb]{.973,.796,.678} 59.4 & \cellcolor[rgb]{.973,.796,.678}72.7  & \cellcolor[rgb]{.957,.69,.518}\textbf{64.1}  \\ \bottomrule[1.5pt]
\end{tabular}
}
\caption{The \textit{test accuracy} (\%) and \textit{clean accuracy} (\%) results on ImageNet dataset using ResNet18 and DenseNet121.}
\label{tab:tab2}
\vspace{-3mm}
\end{table}

\begin{table}[t]
\centering
\scalebox{0.88}{
\begin{tabular}{c|>{\centering\arraybackslash}m{1.5cm}>{\centering\arraybackslash}m{1.5cm}>{\centering\arraybackslash}m{1.5cm}>{\centering\arraybackslash}m{1.5cm}>{\centering\arraybackslash}m{1.8cm}>{\centering\arraybackslash}m{1.5cm}}
\toprule[1.5pt] 
\cellcolor[rgb]{.95,.95,.95}\textbf{Defense$\downarrow$ Model$\rightarrow$}
& \cellcolor[rgb]{.95,.95,.95}ResNet18 & \cellcolor[rgb]{.95,.95,.95}ResNet50  & \cellcolor[rgb]{.95,.95,.95}VGG16  & \cellcolor[rgb]{.95,.95,.95}VGG19 & \cellcolor[rgb]{.95,.95,.95}DenseNet121 & \cellcolor[rgb]{.95,.95,.95}\textbf{AVG}   \\ \midrule[1.5pt]
w/o     & 94.95  & 94.53   & 93.27   & 93.04  & 93.91   & 93.94 \\ \hdashline
AT      & 89.57  & 89.88 & 88.04  & 86.93 & 89.11  & 88.71        \\
ISS-J   & 85.23 & 85.85  & 84.42  & 84.26 & 85.08 & 84.97   \\
\textbf{ECLIPSE (Ours)} &\cellcolor[rgb]{.957,.69,.518} \textbf{90.43} & \cellcolor[rgb]{.957,.69,.518}\textbf{90.17} & \cellcolor[rgb]{.957,.69,.518}\textbf{88.38} & \cellcolor[rgb]{.957,.69,.518}\textbf{88.58} & \cellcolor[rgb]{.957,.69,.518}\textbf{89.35} & \cellcolor[rgb]{.957,.69,.518}\textbf{89.38}       
\\ \bottomrule[1.5pt]
\end{tabular}
}
\caption{The \textit{clean accuracy} (\%) results on CIFAR-10 dataset with SOTA defense schemes across diverse models.}
\label{tab3}
\vspace{-5mm}
\end{table}

\subsection{Purification Visual Effect}
\begin{figure}[t]
    \centering
    \includegraphics[scale=0.45]{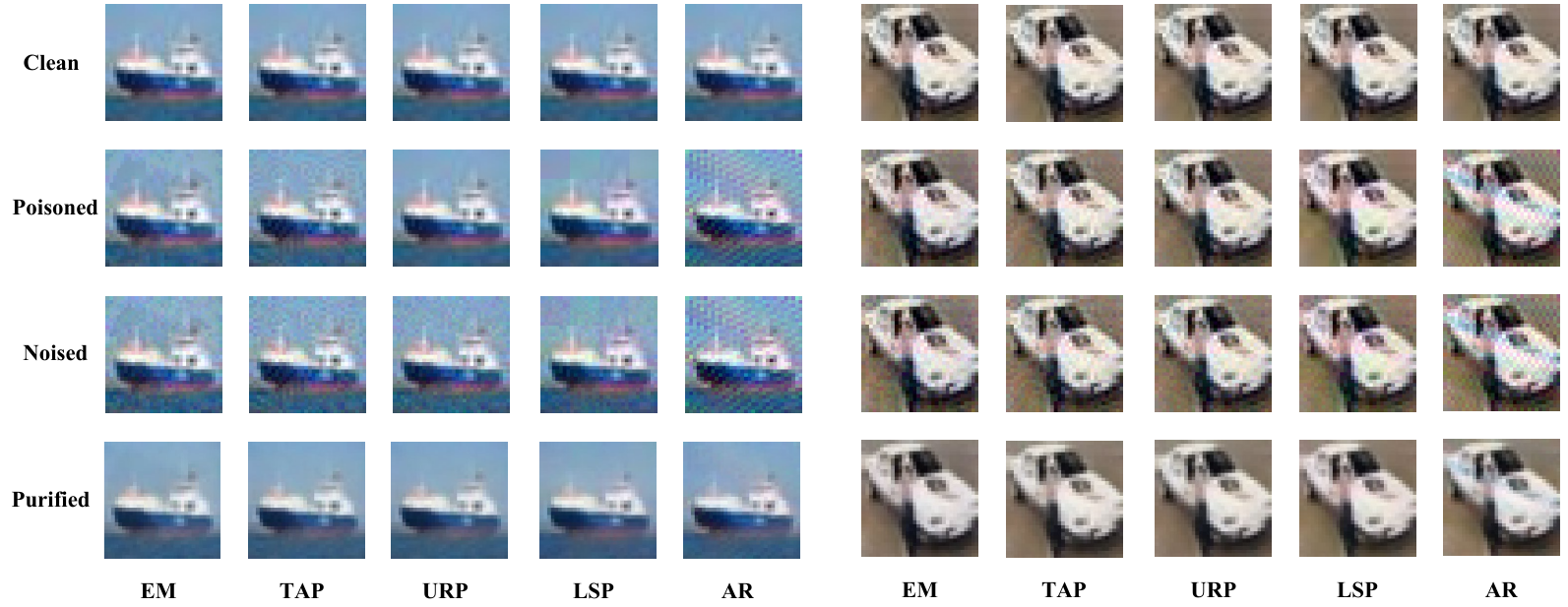}
    \caption{Visual presentations of five types of CLBPAs, including clean, poisoned, noised, and purified images.}
    \label{fig:purify_process}
\end{figure}

After undergoing the processes of poison absorption and noise denoising, the resulting image is essentially a purified image, as demonstrated in~\cref{fig:purify_process}.
It can be seen that the poisoned images clearly have their poison noise removed after passing through our sparse diffusion purification stage.

\subsection{Resistance to Potential Adaptive Attacks}
We assume the attacker has knowledge of the structure of the diffusion model and compensation module, and then design an adaptive attack against ECLIPSE, termed as ADP, which involves solving the following optimization objective:
\begin{equation*}
    \underset{\theta}{\arg \min } \underset{(x, y) \sim \mathcal{D}_{c}}{\mathbb{E}}
    \left[\min _{\boldsymbol{\delta_{a}}} \mathcal{L}\left(U (C(x+\boldsymbol{\delta_{a}});\theta), y\right)\right] \quad 
\end{equation*}
where $\boldsymbol{\left \|\delta_{a}\right \|_\infty \le \epsilon}$ is the adaptive poison with $\ell _{\infty}$-norm bounded, $U$ denotes the U-Net~\cite{zhou2018unet} that forms the diffusion model, which has been slightly modified into a  classification network (the final layer is replaced by a convolutional layer with global average pooling), and $C$ is the corruption function in~\cref{eq13}. The defense performance against ADP is reported in~\cref{tab:tab1}.  Our ECLIPSE demonstrates excellent performance in defending against ADP, indicating the robustness of our defense scheme ECLIPSE when facing the adaptive attack.

\begin{table}[t]
\centering
\scalebox{0.88}{\begin{tabular}{c|>{\centering\arraybackslash}m{1.14cm}>{\centering\arraybackslash}m{1.14cm}>{\centering\arraybackslash}m{1.14cm}>{\centering\arraybackslash}m{1.14cm}>{\centering\arraybackslash}m{1.14cm}>{\centering\arraybackslash}m{1.14cm}>{\centering\arraybackslash}m{1.14cm}>{\centering\arraybackslash}m{1.14cm}>{\centering\arraybackslash}m{1.14cm}}
\toprule[1.5pt]
Module$\downarrow$ Poison$\rightarrow$ & EM    & TAP   & REM   & SEP   & EFP   & URP   & LSP   & AR    & \textbf{AVG}   \\ \hline
\cellcolor[rgb]{.90588,.90196,.90196}A+B+C & \cellcolor[rgb]{.90588,.90196,.90196}82.80 & \cellcolor[rgb]{.90588,.90196,.90196}86.13 & \cellcolor[rgb]{.90588,.90196,.90196}82.72 & \cellcolor[rgb]{.90588,.90196,.90196}82.85 & \cellcolor[rgb]{.90588,.90196,.90196}77.20 & \cellcolor[rgb]{.90588,.90196,.90196}86.98 & \cellcolor[rgb]{.90588,.90196,.90196}84.58 & \cellcolor[rgb]{.90588,.90196,.90196}87.32 & \textbf{\cellcolor[rgb]{.90588,.90196,.90196}83.82} \\
A+B  & 73.22 & 86.31 & 67.48 & 41.30 & 77.58 & 85.59 & 81.58 & 84.22 & 74.66   \\
A+C  & 61.30 & 86.94 & 77.33 & 84.07 & 76.72 & 87.52 & 68.70 & 87.59 & 78.77  \\
B+C  & 78.82 & 83.32 & 75.51 & 14.66 & 81.62 & 88.87 & 80.08 & 60.96 & 70.48  \\ 
A &27.45 &86.63 &35.74 &44.97 &75.90 &86.86 &39.93 	&83.98 	&60.18\\
B &78.69 &30.95 &67.05 &7.53 &86.14 &58.70 &68.19 &37.63 &54.36 \\
C &20.43 	&84.90 	&28.83 	&11.93 	&79.06 	&89.54 	&27.11 	&33.14 	&46.87 \\
\bottomrule[1.5pt]
\end{tabular}}
\caption{The \textit{test accuracy} (\%) results on CIFAR-10 using ResNet18 with diverse combinations. ``A", ``B", ``C" denote  diffusion purification, grayscale module, and Gaussian noise module. The \colorbox[RGB]{231,230,230}{gray line}  denotes the best effect in this paper.}
\label{tab4} 
\vspace{-5mm}
\end{table}

\begin{table}[t]
\centering
\scalebox{0.88}{\begin{tabular}{c|>{\centering\arraybackslash}m{1.16cm}>{\centering\arraybackslash}m{1.16cm}>{\centering\arraybackslash}m{1.16cm}>{\centering\arraybackslash}m{1.16cm}>{\centering\arraybackslash}m{1.16cm}>{\centering\arraybackslash}m{1.16cm}>{\centering\arraybackslash}m{1.16cm}>{\centering\arraybackslash}m{1.16cm}>{\centering\arraybackslash}m{1.16cm}}
\toprule[1.5pt]
$M$$\downarrow$ Poison$\rightarrow$ & EM    & TAP   & REM   & SEP   & EFP   & URP   & LSP   & AR    & \textbf{AVG}   \\ \hline
0 & 78.30 & 86.59 & 81.56 & 77.97 & 78.60 & 87.52 & 83.53 & 87.56 & 82.70          \\
2 & 79.72 & 86.88 & 82.05 & 81.29 & 78.02 & 86.77 & 84.50 & 86.98 & 83.28          \\
\cellcolor[rgb]{.90588,.90196,.90196}4 & \cellcolor[rgb]{.90588,.90196,.90196}82.80 & \cellcolor[rgb]{.90588,.90196,.90196}86.13 & \cellcolor[rgb]{.90588,.90196,.90196}82.72 & \cellcolor[rgb]{.90588,.90196,.90196}82.85 & \cellcolor[rgb]{.90588,.90196,.90196}77.20 & \cellcolor[rgb]{.90588,.90196,.90196}86.98 & \cellcolor[rgb]{.90588,.90196,.90196}84.58 & \cellcolor[rgb]{.90588,.90196,.90196}87.32 & \textbf{\cellcolor[rgb]{.90588,.90196,.90196}83.82} \\
6 & 80.75 & 86.79 & 82.68 & 82.71 & 78.41 & 86.52 & 83.88 & 87.05 & 83.60          
\\ \bottomrule[1.5pt]
\end{tabular}}
\caption{The \textit{test accuracy} (\%) results on CIFAR-10 using ResNet18 with varying replication times $M$ from ECLIPSE.}
\label{tab5}
\vspace{-3mm}
\end{table}

\vspace{-3mm}
\subsection{Hyper-Parameter Analysis}
\begin{figure*}[t]
\centering
\subcaptionbox{$p$=0.4, $\sigma$=0.05}
{\includegraphics[width=0.327\textwidth]{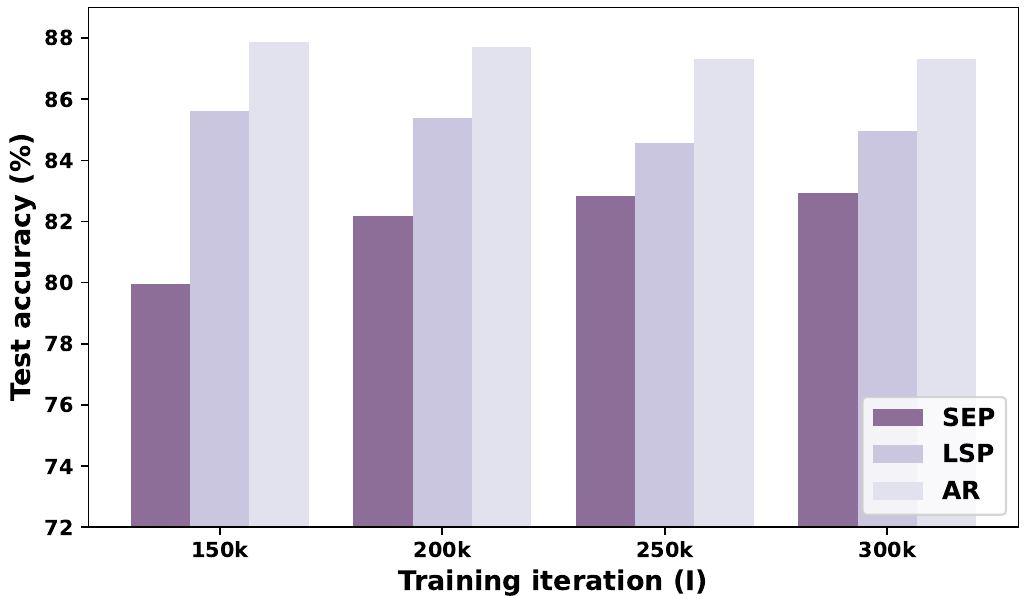}}
\subcaptionbox{$\sigma$=0.1, $I$=200K}{\includegraphics[width=0.327\textwidth]{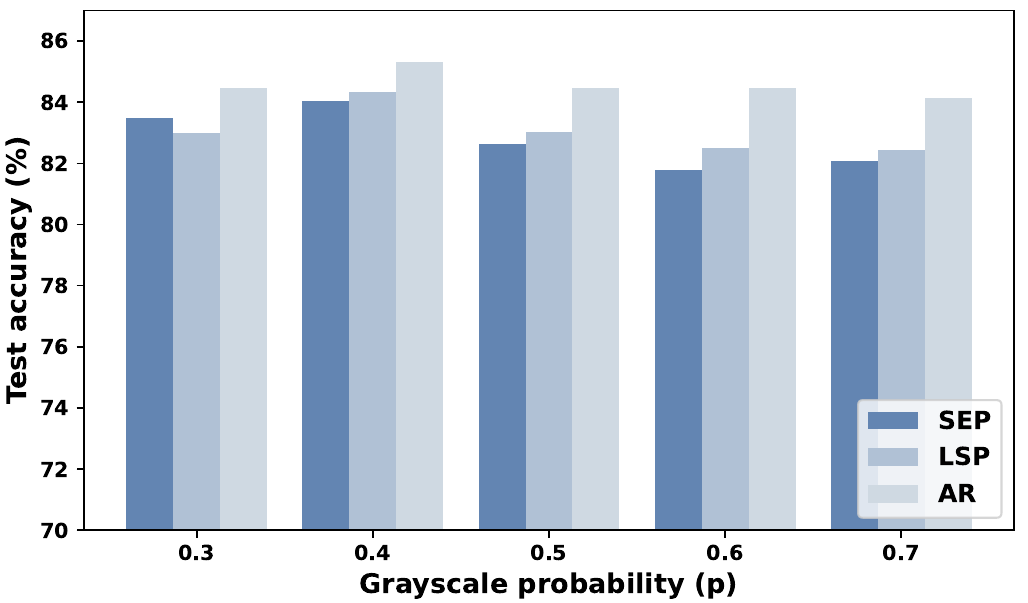}}
\subcaptionbox{$p$=0.4, $I$=200K}{\includegraphics[width=0.329\textwidth]{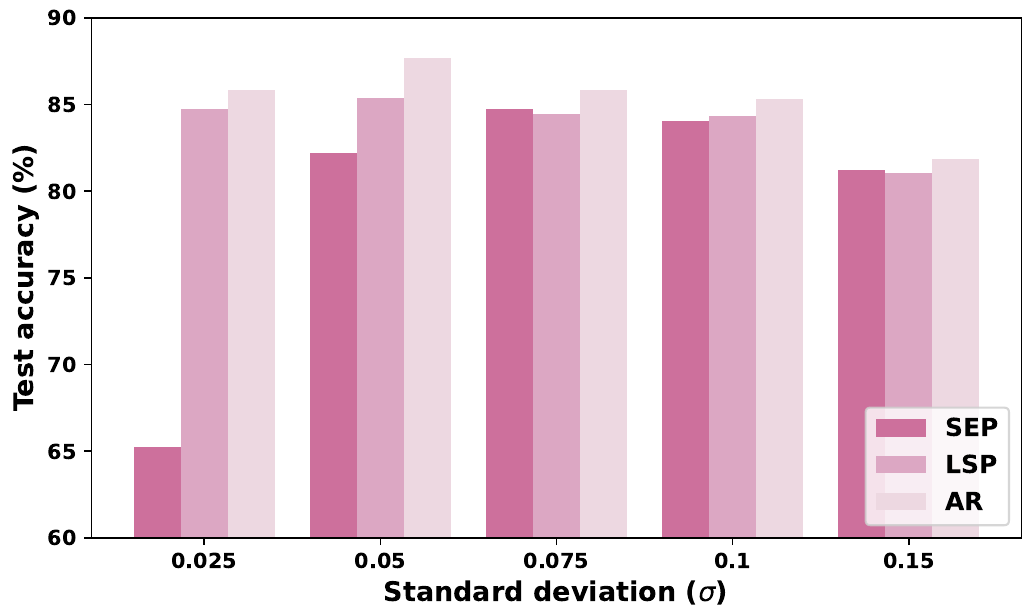}}
 \caption{The \textit{test accuracy} (\%) results of ECLIPSE on three poisoned CIFAR-10 dataset using ResNet18 with varying hyper-parameters.}
 \label{fig:4}
\end{figure*}

We investigate the effects of hyperparameters $I$, $p$,  $\sigma$, $M$, $B$/$N$, and $t^*$ on the ECLIPSE defense performance. 
It can be observed from~\cref{fig:4} (a) that different values of $I$ had little effect on the final defense performance, which can be attributed to the relatively small size of the training set, allowing the diffusion model to converge well within the above range of $I$. 
The impact of different probabilities $p$ of Module B on ECLIPSE can be found in~\cref{fig:4} (b).
The excessively low probability of grayscale transformation may lead to incomplete removal of low-frequency poisons, whereas an excessively high probability may compromise certain image features.  
The impact of different standard deviations $\sigma$ on ECLIPSE is provided in~\cref{fig:4} (c). 
Both excessively large and excessively small Gaussian noise lead to a decline in the final defensive performance. 
In addition, we explore the impact from the replication times $M$ in~\cref{tab5}, the ratio of sparse set ($B/N$) in~\cref{tab6}, and the forward timestep $t^*$ in~\cref{tab7}. We can see that $M$ mainly influences the defense effect against EM and SEP, which is improved by up to 4.5\% and 4.88\% compared to not using dataset enlarging module. 
The potential reason for this  improvement is that expanding the total amount of data might help the diffusion model better learn the existing distribution of clean data.
The proportion of the sparse set ($B/N$) does not significantly influence the defense effect. Setting the value of $t^*$ too large or too small will both lead to a decrease in defense effect.

\subsection{Ablation Study}
\label{sec:ablation}
Our defense can be divided into three modules, diffusion purification, probabilistic grayscale transformation, and adding Gaussian noise.

\noindent\textbf{Diffusion purification (A).}  The results obtained using Module B+C in~\cref{tab4} indicate the absence of Module A leads to a significant decrease in defense performance against SEP and AR, with a 13.34\% average performance drop. This strongly demonstrates the importance of incorporating Module A.

\begin{table}[t]
\centering
\scalebox{0.88}{\begin{tabular}{c|>{\centering\arraybackslash}m{1.16cm}>{\centering\arraybackslash}m{1.16cm}>{\centering\arraybackslash}m{1.16cm}>{\centering\arraybackslash}m{1.16cm}>{\centering\arraybackslash}m{1.16cm}>{\centering\arraybackslash}m{1.16cm}>{\centering\arraybackslash}m{1.16cm}>{\centering\arraybackslash}m{1.16cm}>{\centering\arraybackslash}m{1.16cm}}
\toprule[1.5pt]
$B/N$ $\downarrow$ Poison$\rightarrow$ & EM    & TAP   & REM   & SEP   & EFP   & URP   & LSP   & AR    & \textbf{AVG}   \\ \hline
2\%   & 80.25 & 84.28 & 82.45 & 82.39 & 73.72 & 84.70 & 82.04 & 85.08 & 81.86          \\
\cellcolor[rgb]{.90588,.90196,.90196}4\%                                 & \cellcolor[rgb]{.90588,.90196,.90196}82.80 & \cellcolor[rgb]{.90588,.90196,.90196}86.13 & \cellcolor[rgb]{.90588,.90196,.90196}82.72 & \cellcolor[rgb]{.90588,.90196,.90196}82.85 & \cellcolor[rgb]{.90588,.90196,.90196}77.20 & \cellcolor[rgb]{.90588,.90196,.90196}86.98 & \cellcolor[rgb]{.90588,.90196,.90196}84.58 & \cellcolor[rgb]{.90588,.90196,.90196}87.32 & \cellcolor[rgb]{.90588,.90196,.90196}\textbf{83.82} \\
6\%  & 80.07 & 87.05 & 79.61 & 78.51 & 79.82 & 87.54 & 84.77 & 88.25 & 83.20 \\
8\% & 78.13 & 86.75 & 77.65 & 65.51 & 80.39 & 88.26 & 81.82 & 87.48 & 80.75 \\ \bottomrule[1.5pt]
\end{tabular}}
\caption{The \textit{test accuracy} (\%) results on CIFAR-10 using ResNet18 with varying dataset ratios $B/N$ (\%) from ECLIPSE.}
\label{tab6}
\end{table}

\begin{table}[t]
\centering
\scalebox{0.88}{\begin{tabular}{c|>{\centering\arraybackslash}m{1.16cm}>{\centering\arraybackslash}m{1.16cm}>{\centering\arraybackslash}m{1.16cm}>{\centering\arraybackslash}m{1.16cm}>{\centering\arraybackslash}m{1.16cm}>{\centering\arraybackslash}m{1.16cm}>{\centering\arraybackslash}m{1.16cm}>{\centering\arraybackslash}m{1.16cm}>{\centering\arraybackslash}m{1.16cm}}
\toprule[1.5pt]
$t^*$ $\downarrow$ Poison$\rightarrow$ & EM    & TAP   & REM   & SEP   & EFP   & URP   & LSP   & AR    & \textbf{AVG}   \\ \hline
50                                                                     & 77.23 & 87.66 & 78.20 & 55.07 & 81.33 & 88.43 & 81.97 & 88.27 & 79.77 \\ \cellcolor[rgb]{.90588,.90196,.90196}100  
& \cellcolor[rgb]{.90588,.90196,.90196}82.80 
& \cellcolor[rgb]{.90588,.90196,.90196}86.13 
& \cellcolor[rgb]{.90588,.90196,.90196}82.72 
& \cellcolor[rgb]{.90588,.90196,.90196}82.85 
& \cellcolor[rgb]{.90588,.90196,.90196}77.20 
& \cellcolor[rgb]{.90588,.90196,.90196}86.98 
& \cellcolor[rgb]{.90588,.90196,.90196}84.58 
& \cellcolor[rgb]{.90588,.90196,.90196}87.32 
& \cellcolor[rgb]{.90588,.90196,.90196}\textbf{83.82} 
\\ 150 & 80.91 & 84.30 & 82.56 & 83.99 & 74.23 & 83.25 & 79.46 & 84.37 & 81.63 \\ \bottomrule[1.5pt]
\end{tabular}}
\caption{The \textit{test accuracy} (\%) results on CIFAR-10 using ResNet18 with varying forward timestep $t^*$ from ECLIPSE.}
\label{tab7}
\vspace{-3mm}
\end{table}

\noindent\textbf{Probabilistic grayscale transformation (B).} 
The absence of Module B results in varying degrees of negative effects on low-frequency poisons as shown in~\cref{tab4} using Module A+C. This effectiveness of Module B against these poisons is attributed to the fact that low-frequency noise typically affects the color channel information of the images, while grayscale transformation can mask color information and thus suppress low-frequency poisons. As for the reason Module A and Module C are less effective against low-frequency poisons is they rely on adding high-frequency Gaussian noise, thus absorbing low-frequency poisons at a slower rate.

\begin{figure}[t]
    \centering
    \includegraphics[scale=0.65]{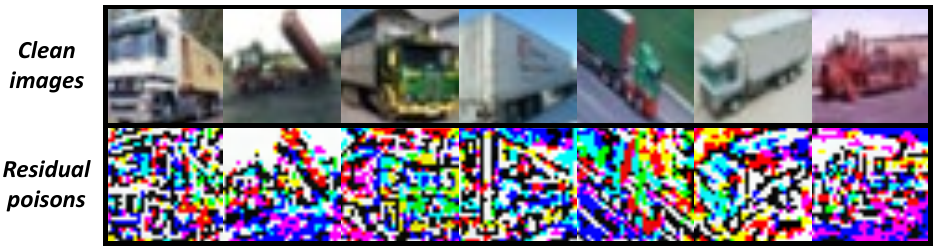}
    \caption{The residual poisons refer to difference between the SEP images after Module A and the corresponding clean images, which shows that the images in two rows exhibit a high feature similarity.}
    \label{fig5} 
    \vspace{-3mm}
\end{figure}

\noindent\textbf{Addition of Gaussian noise (C).} 
The absence of Module C results in a significant decline of 41.55\% against SEP when using Module A+B as shown in~\cref{tab4}. To analyze this, we visualize the remaining SEP perturbations after passing through Module A in~\cref{fig5}. 
Surprisingly, the remaining perturbations are highly similar to the clean samples' features. 
These high-frequency residual poisons with significant feature similarity can be easily learned by DNNs as shortcuts~\cite{shortcut,hu2023pointcrt,liu2023detecting}, leading to a decline in test accuracy. 
Consequently, by introducing Module C that adds random noise, these relatively fragile high-frequency poisons can be effectively disturbed.

\subsection{Analysis of ECLIPSE}
\label{sec:analysis}
We categorize existing CLBPAs into three types: low-frequency poisons (\ie, EM, REM, EFP, and LSP), robust high-frequency poisons (\ie, SEP), and fragile high-frequency poisons (\ie, TAP, URP, and AR). 
We conclude that,  
\textbf{(i)} \textit{Module A can remove high-frequency poisons.} The effectiveness of Module A against fragile high-frequency poisons can be verified in~\cref{tab4}. In addition, by increasing $t^*$, Module A can effectively remove robust high-frequency poisons as shown in~\cref{tab7}, 
\textbf{(ii)}  \textit{Module B is crucial for addressing low-frequency poisons.} It is evident from~\cref{tab4} that as long as Module B is present (\ie, A+B+C, A+B, B+C, and B), the defense against low-frequency poisons is highly effective, and 
\textbf{(iii)} \textit{Module C has a positive impact in expunging low-frequency, robust high-frequency, and fragile high-frequency poisons.} 
The incremental effect brought by Module C can be observed by comparing the results before and after its inclusion as demonstrated in~\cref{tab4}.  Thus, lightweight Gaussian noise is indeed beneficial for expunging CLBPA poisons.

\section{Conclusion and Limitation}
We propose a brand-new defense scheme called ECLIPSE against the recent rise of clean-label indiscriminate poisoning attacks, which is more universally effective, more practical, and robust. Our scheme is capable of expunging diverse invisible poisons of images via a purification process by a sparse data trained diffusion model and a compensation module. Extensive experiments on benchmark datasets verify that our scheme enjoys high defense effectiveness and robustness against adaptive attack. However, 
ECLIPSE is not that effective when defending against clean-label indiscriminate poisoning attacks based on convolution-based perturbations~\cite{sadasivan2023cuda,wang2023corrupting} without norm restriction. 
This is because the purification scheme based on diffusion models can only remove perturbations with norm constraints~\cite{diffpure}. 
We will leave this issue to our future work.

\section*{Acknowledgements}   
This work is supported by the National Natural Science Foundation of China (Grant No.U20A20177) and Hubei Province Key R\&D Technology Special Innovation Project (Grant No.2021BAA032). Shengshan Hu and Peng Xu are co-corresponding authors.

\section*{Appendix}
\label{sec:appendix}
\noindent\textbf{Proof for Theorem 1:} 
Based on the continuous-time forward process defined as the solution to the SDE~\cite{song2021maximum}, we have:
\begin{align}
    \mathrm{d}x=g(t)\mathrm{d}\beta +  h(x,t)\mathrm{d}t
\end{align}
where $g(t)$ is the diffusion coefficient, $h(x,t)$ is the drift coefficient, and $\beta(t)$ is a Brownian motion with a diffusion matrix. After this, according to the Fokker-Planck-Kolmogorov equation~\cite{sarkka2019applied}, we have:
{
\small
\begin{align}
    \frac{\partial p(x,t)}{\partial t}&= -\nabla_x\left(h(x,t)p(x,t)-\frac{g^2(t)}{2}\nabla_xp(x,t) \right) 
    \notag\\&=-\nabla_x\left(h(x,t)p(x,t)-\frac{g^2(t)}{2}p(x,t)\nabla_x \log_{} {p(x,t)} \right)
    \notag\\&=\nabla_x\left(k_p(x,t)p(x,t) \right)
\end{align}
}
where $k_p(x,t)$ is defined as $-h(x,t)+\frac{\nabla_x \log_{} {p(x,t)}}{2}g^2(t)$.
Then we have:
{
\small
\begin{align}
    \frac{\partial  D_{KL}\left(p(x, t) \| q(x, t)\right)}{\partial t}&=
    \frac{\partial }{\partial t } \int p(x,t)log\frac{p(x,t)}{q(x,t)}dx
   \notag\\&=\int \frac{\partial p(x,t)}{\partial t} \log_{}{\frac{p(x,t)}{q(x,t)}dx} +\int \frac{p(x,t)}{q(x,t)}\frac{\partial q(x,t)}{\partial t}dx
    +  \int \frac{\partial p(x,t)}{\partial t}dx
    \notag\\&=\int \nabla_x\left(k_p(x,t)p(x,t) \right) \log_{}{\frac{p(x,t)}{q(x,t)}dx} + \int \nabla_x\left(k_q(x,t)q(x,t) \right)\frac{p(x,t)}{q(x,t)}dx
    \notag\\&=
    -\int p(x,t)\left [ k_p(x,t)-k_q(x,t) \right ] ^T[\nabla _x\log_{}{p(x,t)}
    -\nabla _x\log_{}{q(x,t)} ]dx
    \notag\\&=-\frac{g^2(t)}{2} \int p(x,t)\left \| \nabla _x\log_{}{p(x,t)}-\nabla _x\log_{}{q(x,t)}  \right \|_{2}^{2}dx  
    \notag\\&=-\frac{g^2(t)}{2} D_F(p(x,t)||q(x,t))
\end{align}
}where the fourth equality follows from the integration by parts and our assumption of smooth and fast-decaying $p(x,t)$ and $q(x,t)$. Here, $D_F$ denotes the Fisher divergence~\cite{kostrikov2021offline}. 
Since $g^2(t) > 0$ and the Fisher divergence is non-negative, we have:
\begin{align}
    \frac{\partial  D_{KL}\left(p(x, t) \| q(x, t)\right)}{\partial t} \le 0
\end{align}
where equality holds only if $p(x,t)=q(x,t)$.

\bibliographystyle{splncs04}
\bibliography{ref}
 
\end{document}